\documentclass[proceedings, preprint]{rmaa}

\usepackage{paralist}

\usepackage{psfrag,color}
\usepackage{epstopdf}

\SetYear{2013}
\SetConfTitle{Magnetic Fields in the Universe IV} 

\title{Magnetic Fields of Neutron Stars}

\author{
  Andreas Reisenegger\altaffilmark{1}
  }

\altaffiltext{1}{Departamento de Astronom{\'\i}a y Astrof{\'\i}sica,
Pontificia Universidad Cat\'olica de Chile, Av. Vicu\~na Mackenna 4860, 7820436 Macul, Santiago, Chile (areisene@astro.puc.cl).}

\shortauthor{Reisenegger}
\shorttitle{Magnetic fields of neutron stars}

\listofauthors{A. Reisenegger}
\indexauthor{Reisenegger, A.}

\abstract{Neutron stars contain the strongest magnetic fields known in the Universe. In this paper, I discuss briefly how these magnetic fields are inferred from observations, as well as the evidence for their time-evolution. I show how these extremely strong fields are actually weak in terms of their effects on the stellar structure, as is also the case for magnetic stars on the upper main sequence and magnetic white dwarfs, which have similar total magnetic fluxes. I propose a scenario in which a stable hydromagnetic equilibrium (containing a poloidal and a toroidal field component) is established soon after the birth of the neutron star, aided by the strong compositional stratification of neutron star matter, and this state is slowly eroded by non-ideal magnetohydrodynamic processes such as beta decays and ambipolar diffusion in the core of the star and Hall drift and breaking of the solid in its crust. Over sufficiently long time scales, the fluid in the neutron star core will behave as if it were barotropic, because, depending on temperature and magnetic field strength, beta decays will keep adjusting the composition to the chemical equilibrium state, or ambipolar diffusion will decouple the charged component from the neutrons. Therefore, the still open question regarding stable hydromagnetic equilibria in barotropic fluids will become relevant for the evolution, at least for ``magnetar'' fields, too strong to be stabilized by the solid crust.}

\resumen{Las estrellas de neutrones contienen los campos magn\'eticos m\'as intensos conocidos en el Universo. En este art{\'\i}culo, discuto brevemente c\'omo estos campos son inferidos de las observaciones, as{\'\i} como la evidencia de su evoluci\'on temporal. Muestro que estos campos extremadamente intensos son en realidad d\'ebiles en cuanto a sus efectos sobre la estructura estelar, como tambi\'en es el caso para las estrellas magn\'eticas en la secuencia principal superior y las enanas blancas magn\'eticas. Propongo un escencario en el cual poco despu\'es del nacimiento de la estrella de neutrones se establece un equilibrio hidromagn\'etico estable (en que el campo magn\'etico tiene componentes poloidales y toroidales), con ayuda de la fuerte estratificaci\'on composicional de la materia, y este estado es erosionado paulatinamente por procesos de magnetohidrodin\'amica no ideal, como decaimientos beta y difusi\'on ambipolar en el n\'ucleo de la estrella, y deriva de Hall y rompimiento del s\'olido en la corteza. En tiempos suficientemente prolongados, el fluido del n\'ucleo se comportar\'a como si fuera barotr\'opico, porque, dependiendo de la intensidad del campo y de la temperatura, decaimientos beta ir\'an ajustando la composici\'on de la materia al correspondiente equilibrio qu\'imico, o la difusi\'on ambipolar desacoplar\'a a la componente cargada de los neutrones. Por esta raz\'on, la pregunta a\'un abierta acerca de la existencia de equilibrios hidromagn\'eticos estables en fluidos barotr\'opicos ser\'a relevante para la evoluci\'on, por lo menos para los campos de magnetares, demasiado intensos para ser estabilizados por la corteza s\'olida.}

\addkeyword{stars: neutron}
\addkeyword{stars: magnetic field}
\addkeyword{stars: magnetars}
\addkeyword{(stars:) pulsars: general}

\begin{document}
\maketitle

\section{Basic theory: Structure and composition}
\label{sec:theory}

Neutron stars are very compact stellar remnants, whose extremely strong gravity is balanced by the gradient of the pressure of highly degenerate fermions, mostly neutrons. Moving inside from the neutron star surface, one can distinguish qualitatively different layers of matter:
\begin{itemize}
\item The \emph{outer crust} (densities $\rho\sim 10^6-4\times 10^{11}\mathrm{g\,cm^{-3}}$), a solid of heavy nuclei and freely moving electrons.
\item The \emph{inner crust} ($\rho\sim 4\times 10^{11}-2\times 10^{14}\mathrm{g\,cm^{-3}}$), a solid of even heavier nuclei, freely moving electrons, and freely moving neutrons, the latter likely in a superfluid state.
\item The \emph{outer core} ($\rho\sim 2\times 10^{14}-10^{15}\mathrm{g\,cm^{-3}}$), a liquid composed mostly of neutrons ($n$), with a relatively small fraction (few $\%$) of protons ($p$), electrons ($e$), and muons.
\item The \emph{inner core} ($\rho\gtrsim 10^{15}\mathrm{g\,cm^{-3}}$), in a largely unknown state, likely a liquid containing more exotic particles, such as mesons, hyperons, free quarks, or others.
\end{itemize}

The presence of (at least) protons and electrons in the outer core is crucial, because they block quantum states into which the neutrons would otherwise decay, as they do (with a half-life of only 15 minutes) in the relative vacuum of our own surroundings. Beta equilibrium, in which the reaction $n\to p+e+\bar\nu_e$ is in balance with its counterpart $p+e\to n+\nu_e$, is set by the condition $\mu_n=\mu_p+\mu_e$, where $\mu_i$ is the chemical potential (the Fermi energy, corrected by strong interactions and very small thermal effects) of particle species $i$. This condition requires the fraction of protons and electrons (compared to neutrons) to be an increasing function of density. This means that the fluid is \emph{stably stratified}, resisting convective turnover, like water with downward-increasing salinity (Pethick 1991; Reisenegger \& Goldreich 1992).

Protons and electrons are also important as charge carriers, which allow currents to flow, and thus a magnetic field to be supported inside the neutron star. Since these particles are highly degenerate, most of their quantum states are occupied, and thus it is difficult to scatter them into a different state. For this reason, the resistivity is low, and currents can flow for a long time without being dissipated, even more so if, as expected, the protons in much of the core are superconducting (Baym, Pethick, and Pines 1969a,b).

\section{Basic observations: Spin-down, magnetic field, and classes of neutron stars}
\label{sec:observations}

The electromagnetic radiation received from most neutron stars appears pulsed at a very regular frequency, which slowly decreases in time. This is almost certainly due to the slowing rotation rate $\Omega$ of the neutron star, whose radiation is beamed or at least anisotropic. The slow-down is usually modeled (not quite realistically) in terms of a magnetic dipole rotating in vacuum, which loses rotational energy through electromagnetic radiation according to the relation
\begin{equation}
I\Omega\dot\Omega\propto-\mu^2\Omega^4,   \label{eq:spin-down}
\end{equation}
where dots indicate time derivatives, $I$ is the moment of inertia, and $\mu$ is the magnetic moment of the star. This allows to estimate the spin-down time, $t_s\equiv P/(2\dot P)$, as a rough estimate of the stellar age (accurate if $\mu=$ constant and the initial rotation rate was much faster than the present one), and the surface magnetic field $B\propto(P\dot P)^{1/2}$, where $P=2\pi/\Omega$ is the rotation period.

In nearly all cases, there are no other measurements of the magnetic field strength, and only indirect inferences of its geometry from the pulse profiles. However, the magnetic field clearly plays an important role in neutron star evolution and is present on all known neutron stars. Its magnitude is inferred to be $10^{11-13}\mathrm{G}$ in most objects (the bulk of the so-called ``classical pulsars''), as low as $10^{8-9}\mathrm{G}$ in the old, but rapidly spinning ``millisecond pulsars'', and as high as $10^{14-15}\mathrm{G}$ in the slowly spinning ($P\sim 2-12\mathrm{s}$), but very energetic ``soft gamma repeaters'' (SGRs) and ``anomalous X-ray pulsars'' (AXPs), collectively known as ``magnetars''. In addition to these, one phenomenologically distinguishes isolated thermal emitters (INSs; $B\sim 10^{13-14}\mathrm{G}$), ``central compact objects'' in supernova remnants (CCOs; $B\sim 10^{10-12}\mathrm{G}$), RRATs (intermittent radio pulsars; $B\sim 10^{12-14}\mathrm{G}$), and accreting neutron stars (high-mass and low-mass X-ray binaries). A concise overview of these classes of neutron stars, their position on the $P-\dot P$ diagram, and their possible connections is given by Kaspi (2010).

There are several lines of evidence suggesting possible evolution of the magnetic field:
\begin{itemize}
\item Field decay inferred from the distribution of classical pulsars on the $P-\dot P$ diagram: complicated by a number of selection effects, it has been addressed by many authors over the last 35 years, with conflicting results. For a recent analysis, see Faucher-Gigu{\`e}re \& Kaspi (2006).

\item Very weak (dipole) field of old, recycled pulsars (millisecond pulsars and low-mass X-ray binaries): not yet established whether this is an effect of age (passive magnetic field decay) or induced by accretion (increased resistivity due to heating, magnetic field burial, or motion of superfluid neutron vortices).

\item Anomalous braking indices: In very young pulsars, it is possible to measure $\ddot\Omega$ and thus construct the ``braking index'' $n\equiv\Omega\ddot\Omega/\dot\Omega^2$. Eq.~(\ref{eq:spin-down}) with $\mu=$ constant yields $n=3$, whereas measured values are generally lower, at face value implying an increasing magnetic dipole moment.

\item Magnetar energetics: SGRs and AXPs emit copious amounts of high-energy (X and gamma) radiation; in fact, their time-averaged bolometric luminosity exceeds the rotational energy loss given by eq.~(\ref{eq:spin-down}). This suggested that their energy source might be the decay of their magnetic field (Thompson \& Duncan 1996), later corroborated by the determination of their dipole field as the highest known for any objects (Kouveliotou et al. 1998). Note, however, that an even stronger internal field appears to be required to account for the energetics of some of these objects. An interesting, recent discovery has been the detection of quasi-periodic oscillations following SGR flares (Israel et al. 2005), which might be magneto-elastic oscillation modes of the neutron star and thus potential probes of its internal magnetic field structure.
\end{itemize}

\section{Field strength in context: Very strong \emph{and} very weak}
\label{sec:strength}

As already mentioned, the observationally inferred dipole magnetic fields of neutron stars, particularly magnetars ($B\sim 10^{14-15}\mathrm{G}$), are the strongest known in the Universe, far exceeding any produced so far on Earth (up to $10^7\mathrm{G}$ produced in explosions, for very short times) or on other stars (up to $10^9\mathrm{G}$ on white dwarfs). An interesting comparison table is given on R. Duncan's web site on magnetars (\url{http://solomon.as.utexas.edu/magnetar.html}).

\begin{table*}[!t]\centering
  \newcommand{\DS}{\hspace{6\tabcolsep}} 
  \setlength{\tabnotewidth}{0.9\textwidth}
  \setlength{\tabcolsep}{1.33\tabcolsep}
  \tablecols{4}
  \caption{Stars with long-lived magnetic fields} \label{tab:stars}
  \begin{tabular}{lccc}
    \toprule
    Star type & \multicolumn{1}{c}{Upper main sequence} & \multicolumn{1}{c}{White dwarf} & \multicolumn{1}{c}{Neutron star} \\
    \midrule
    Radius $R$ [km]    & $10^{6.5}$ & $10^4$ & $10^1$\\
    Maximum magnetic field $B_{max}$ [G]    & $10^{4.5}$ & $10^9$ & $10^{15}$\\
    Maximum magnetic flux $\Phi_{max}\equiv\pi R^2 B_{max}$ [$\mathrm{G\,km^2}$] & $10^{18}$ & $10^{17.5}$ & $10^{17.5}$\\
    \bottomrule
  \end{tabular}
\end{table*}

On the other hand, neutron stars share with white dwarfs and upper main sequence stars the properties of being mostly or completely non-convecting and having fields appearing to be constant over long time scales and thus likely ``frozen in'' rather than being rearranged and regenerated by a dynamo process. Table \ref{tab:stars} shows that the widely different sizes and observed magnetic field strengths among these three types of stars largely compensate to give quite similar maximum magnetic fluxes $\Phi_{max}\sim 10^{17.5-18}\mathrm{G\,km^2}$ in each type, possibly indicating that the naive hypothesis of flux freezing along the evolution of these stars goes a long way in explaining their magnetic fluxes, despite their very eventful lifes, including core collapse, ejection of a substantial fraction of their mass, differential rotation, and convection.

It is interesting to consider the ratio of gravitational to magnetic energy in these stars,
\begin{equation} \label{eq:ratio}
{|E_{grav}|\over E_{mag}}\sim{GM^2/R\over B^2R^3/6}\sim 6\pi^2G\left(M\over\Phi\right)^2 \gtrsim  10^6
\end{equation}
which remains constant as the star contracts or expands, as long as it conserves its mass and magnetic flux. The lower bound, based on the numbers in Table~\ref{tab:stars}, shows that all these stars are very highly ``supercritical'' (in star-formation jargon), so the magnetic forces are much too weak to significantly affect the stellar structure. In this sense, although magnetar fields are the strongest observed in the Universe, they are still very weak in terms of their effect on the stellar structure. Of course, this ignores an eventual additional field component possibly hidden within the star, mentioned in the previous section, to which I will come back below.

\section{Ideal MHD equilibria with axial symmetry}
\label{sec:MHD}

For the reasons just exposed, it is almost certainly an excellent approximation to write the physical variables characterizing the stellar fluid as the sum of a non-magnetized ``background'' plus a much smaller ``magnetic perturbation'', i.~e., density $\rho=\rho_0+\rho_1$, or pressure $P=P_0+P_1$, where $|\rho_1|/\rho_0\sim|P_1|/P_0\sim B^2/(8\pi P_0)\lesssim 10^{-6}$, according to the estimate of eq.~(\ref{eq:ratio}). In the absence of rotation, the background quantities are spherically symmetric and satisfy the usual hydrostatic equilibrium relation,
\begin{equation} \label{eq:background}
{dP_0\over dr}+\rho_0{d\Psi\over dr}=0,
\end{equation}
where $r$ is the radial coordinate, and $\Psi(r)$ is the gravitational potential, whose magnetic perturbation I ignore for simplicity (``Cowling approximation''). (In this section, I also ignore the shear forces in the solid crust and possible superconducting components in the neutron star core.) On the other hand, since the magnetic field $\vec B(\vec r)$ cannot be spherically symmetric, the hydromagnetic equilibrium equation for the perturbed quantities is generally a three-component vector equation:
\begin{equation} \label{eq:perturb}
\nabla P_1+\rho_1\nabla\Psi={1\over c}\vec j\times\vec B,
\end{equation}
where $c$ is the speed of light and $\vec j=(c/4\pi)\nabla\times\vec B$ is the current density.

As explained in \S~\ref{sec:theory}, neutron star matter is chemically inhomogeneous, characterized by at least one composition variable $Y$, such as the ratio of the proton to neutron density, which beta decays adjust to an equilibrium value over very long time scales, but which will be an independent, conserved quantity over dynamical times. If we assume, for now, a single fluid whose composition is frozen in each fluid element, $P_1$ and $\rho_1$ above can be considered as independent variables that separately adjust to satisfy the hydromagnetic equilibrium equation (\ref{eq:perturb}). Of course, two variables are generally not enough to satisfy three scalar equations, so not every magnetic field structure can be realized as a hydromagnetic equilibrium.

The constraint on the magnetic field structure becomes clearest in axial symmetry, in which the magnetic field must take the form
\begin{equation} \label{eq:B}
\vec B=\nabla\alpha(r,\theta)\times\nabla\phi+\beta(r,\theta)\nabla\phi,
\end{equation}
where $\alpha$ and $\beta$ are (up to this point) arbitrary functions of the spherical coordinates $r$ and $\theta$ (but independent of the azimuthal angle $\phi$, for which I also used $\nabla\phi=\hat\phi/[r\sin\theta]$), and $\nabla\cdot\vec B=0$ is automatically satisfied. In this case, $P_1$ and $\rho_1$ must clearly also depend only on $r$ and $\theta$, so the $\phi$-component of the left-hand side of eq.~(\ref{eq:perturb}) must be identically zero, imposing the same on the right-hand side:
\begin{equation} \label{eq:constraint}
0={1\over c}(\vec j\times\vec B)_\phi={\nabla\beta\times\nabla\alpha\over 4\pi r^2\sin^2\theta},
\end{equation}
thus the gradients $\nabla\alpha$ and $\nabla\beta$ must be parallel everywhere, and $\beta$ must be (at least piece-wise) a function of $\alpha$, $\beta(r,\theta)=\beta[\alpha(r,\theta)]$ (Chandrasekhar \& Prendergast 1956; Mestel 1956). Once this is imposed, only two non-trivial components of eq.~(\ref{eq:perturb}) remain, and these can generally be satisfied by an appropriate choice of the two independent variables $P_1$ and $\rho_1$, as shown for a particular case in Mastrano et al. (2011). Thus, no further constraints need to be imposed on the magnetic field to obtain an MHD equilibrium.

\begin{figure}
\begin{center}
\includegraphics[width=6cm]{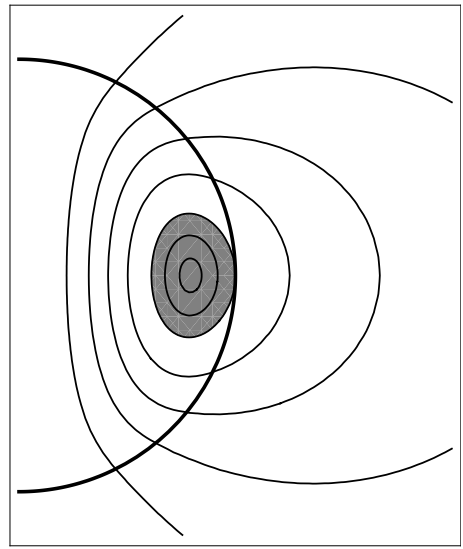}
\caption{Meridional cut of a star with an axially symmetric magnetic field.
The bold curve is the surface of the star, while the thinner curves are poloidal
field lines (corresponding to $\alpha=$ constant). The toroidal component of the
magnetic field ($\beta\neq 0$) lies only in regions where the poloidal field lines
close inside the star (gray region). (Figure prepared by C. Armaza.)}\label{fig:mustlie}
\end{center}
\end{figure}

Fig.~\ref{fig:mustlie} shows what might be an axially symmetric approximation to a realistic magnetic field configuration in a fluid star. The lines shown are the poloidal (meridional) magnetic field lines, i.e., lines of constant $\alpha$. Outside the star, no substantial currents can be present, and this forces the field to be purely poloidal ($\beta=0$). Since $\beta$ is a function of $\alpha$, we will have $\beta=0$ everywhere, except on the field lines that close within the star, corresponding to the shaded region on the plot. In this shaded region, both $\alpha$ and $\beta$ can be non-zero, so the magnetic field lines winds around in a twisted torus, whereas elsewhere $\beta=0$ but $\alpha\neq 0$, so the field lines are purely poloidal, lying in meridional planes.

Long ago, Tayler (1973) showed that, in a stably stratified star, purely toroidal magnetic fields are subject to a kink-type instability, in which flux loops slide with respect to each other, almost exactly on surfaces of constant $r$. Much more recently, Akg\"un et al. (2013) showed that this is true for all toroidal fields, including those confined in a torus, as in Fig.~\ref{fig:mustlie}, which had not been covered by the conditions imposed by Tayler (1973) or in other previous studies. Similarly, it has long been argued that purely poloidal fields ($\alpha\neq 0$, $\beta=0$) are also always unstable (Markey \& Tayler 1973; Wright 1973; Flowers \& Ruderman 1977; Marchant et al. 2011). On the other hand, it was suspected that combined poloidal and toroidal fields might stabilize each other by tying each other together in a magnetic knot as in Fig.~\ref{fig:mustlie}. This appears to be confirmed by the MHD simulations of Braithwaite and collaborators (Braithwaite \& Spruit 2004, 2006; Braithwaite \& Nordlund 2006), in which initially complex magnetic fields generically evolve into nearly axially symmetric, twisted-torus configurations like that in Fig.~\ref{fig:mustlie}, with poloidal and toroidal components of roughly comparable strengths.

The conditions required for the poloidal and toroidal components to stabilize each other were studied numerically by Braithwaite (2009), whereas our group has done a couple of partial, analytical studies (Marchant et al. 2011; Akg\"un et al. 2013). Since the energy in the poloidal field component, $E_{pol}$, can be known or at least estimated from observations (roughly corresponding to $E_{mag}$ in eq.~\ref{eq:ratio}), it is interesting to write the stability conditions as a (very rough and still not rigorously proven) allowed range for the energy in the hidden, toroidal component, $E_{tor}$:
\begin{equation} \label{eq:stability}
0.25 \lesssim {E_{tor}\over E_{pol}} \lesssim 0.5\left[\left({\Gamma\over\gamma}-1\right){|E_{grav}|\over E_{pol}}\right]^{1/2}.
\end{equation}
The indices $\gamma$ and $\Gamma$ characterize, respectively, the equilibrium profile of the star, $\gamma\equiv d\ln P_0/d\ln\rho_0$, and an adiabatic perturbation, which conserves entropy and chemical composition, $\Gamma\equiv(\partial\ln P/\partial\ln\rho)_{ad}$. The Ledoux criterion for stable stratification (stability against convection) requires $\Gamma>\gamma$. In the often assumed barotropic case, $\Gamma=\gamma$, whereas realistic values are $\Gamma/\gamma-1\sim 10^{-2}$ for neutron stars (stabilized by a small fraction of chemical impurities, as discussed in \S~\ref{sec:theory}) and $\Gamma/\gamma-1\sim 1/4$ in the radiative envelopes of upper main sequence stars, which are stabilized by entropy (see Reisenegger 2009 for a more detailed discussion).

Taken at face value, eq.~(\ref{eq:stability}) implies that, for barotropic stars ($\Gamma=\gamma$), there are no (axially symmetric) stable magnetic fields. However, it is important to note that the stars in Braithwaite's simulations were strongly stratified by entropy, whereas the analysis of Akg\"un et al. (2013) assumed strong stable stratification and made approximations based on this assumption. Thus, strictly speaking, neither of them is applicable to the barotropic case. On the other hand, simulations by Lander \& Jones (2012) also suggest that magnetic fields in barotropic stars are generally unstable, and therefore eq.~(\ref{eq:stability}) might be applicable even in that limit.

It is interesting to rewrite the upper limit on $E_{tor}$ from eq.~(\ref{eq:stability}) and evaluate it for neutron stars, in the form
\begin{equation}
{E_{tor}\over|E_{grav}|}\lesssim 0.5\left[\left({\Gamma\over\gamma}-1\right){E_{pol}\over|E_{grav}|}\right]^{1/2}\lesssim 0.5\times 10^{-4},
\end{equation}
where eq.~(\ref{eq:ratio}) was used in the second inequality, identifying $E_{mag}$ there with $E_{pol}$ here. This shows that, for realistic poloidal fields, the toroidal component might be substantially stronger, but the total magnetic energy will still be much smaller than $|E_{grav}|$, so even the toroidal field is weak in a dynamical or structural sense.

\section{Beyond ideal MHD: Magnetic field evolution in neutron stars}
\label{sec:evolution}

In the previous section, I have assumed ideal MHD, in the sense that there is a single, conducting fluid interacting with the magnetic field. This is likely a good approximation in the very early stages of the life of a neutron star, in which the relevant time scales are short and the temperature is high. Initially, the gravitational collapse probably leaves a highly convective, differentially rotating proto-neutron star, which eventually settles into a stable MHD equilibrium like those just described, in just a few Alv\'en times, $t_A\sim R(4\pi\rho)^{1/2}/B\sim(10^{14}\mathrm{G}/B)\mathrm{s}$. Soon afterwards, the temperature decreases enough for the crust to freeze to a solid state, the neutrons of the core and inner crust to become superfluid, and the protons in at least parts of the core to become superconducting.

The crust will thus no longer behave as a fluid. However, the electron currents supporting the magnetic field in the crust will carry along the magnetic flux lines in a process called \emph{Hall drift}, which is non-dissipative but non-linear, possibly leading to a Kolmogoroff-like turbulent cascade of energy to small scales (Goldreich \& Reisenegger 1992) or at least to the formation of current sheets (Urpin \& Shalybkov 1991; Vainshtein et al. 2000; Reisenegger et al. 2007), which dissipate more quickly than a smooth, large-scale current. The evolution of the magnetic field is likely to generate a Lorentz force that can no longer be balanced by pressure and gravity as in eq.~(\ref{eq:perturb}) and will thus produce shear stresses and strains in the solid, which can break the crust if strong enough (as likely in magnetars), causing the matter and the magnetic field to rearrange. How this occurs and whether this can explain some of the violent events in magnetars is still largely an open question (see Levin \& Lyutikov 2012 for a recent discussion).

The core, on the hand, is thought to remain in a fluid state, but here things change as well.

At high temperatures (corresponding to the ``strong-coupling'' regime in the one-dimensional simulations of Hoyos et al. 2008, 2010), the main change is that, over long enough times, neutrons and charged particles can convert into each other through \emph{beta decays}, eventually establishing a chemical equilibrium controlled by only one variable, e.g., the local pressure or density. This means that, in its secular evolution, the fluid will behave as if it were barotropic, with $P_1$ and $\rho_1$ in eq.~(\ref{eq:perturb}) proportional to each other, so there is now only one fluid degree of freedom. In this barotropic state, the possible magnetic field structures are much more constrained, and perhaps no purely fluid, stable equilibria exist. If the field is not too strong, the crust might help in supporting a new equilibrium structure in the core, otherwise the field might break the crust and be largely lost from the star.

At lower temperatures (the ``weak-coupling'' regime of Hoyos et al. 2008, 2010), the fluid becomes more and more degenerate, reducing the phase space for interactions and thus the conversion rates between neutrons and charged particles, but also the drag forces between them, so a two-fluid model becomes more applicable. The magnetic field will be coupled only to the charged particles, and it will force them to move relative to the neutrons in a process called \emph{ambipolar diffusion} (Pethick 1991; Goldreich \& Reisenegger 1992). If the charged particles are only protons and electrons, whose densities are tied together by the condition of charge neutrality, they will behave as a barotropic fluid, bringing us back to the same situation as in the previous paragraph.

\begin{figure}[!ht]
\begin{center} \label{fig:BT}
\includegraphics[width=11.5cm]{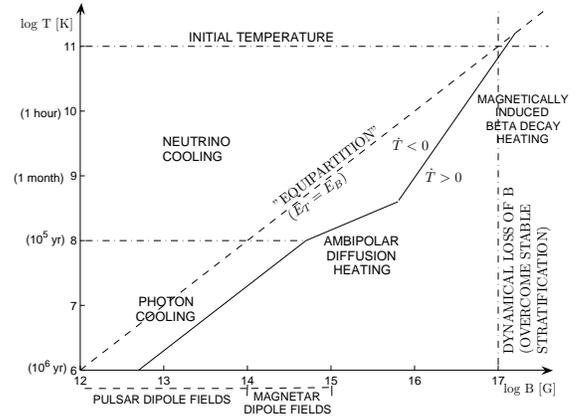}\caption{Magnetic
field -- temperature plane for a non-superfluid neutron star core.
The dot-dashed horizontal lines show the initial temperature (just
after core collapse), and the transition from neutrino-dominated
(modified Urca) to photon-dominated cooling. The dashed diagonal
line corresponds to the equality of magnetic and thermal energy.
Above and to the left of the solid line, the star cools passively,
on the time scales indicated in parenthesis along the vertical
axis, without substantial magnetic field decay, so the evolution
of the star is essentially a downward vertical line. Once the
solid line is reached, magnetic dissipation mechanisms become
important and generate heat that stops the cooling until the magnetic
field has re-arranged to a new equilibrium state. (Figure prepared by
C. Petrovich and first published in Reisenegger 2009.)}
\end{center}
\end{figure}

Thus, the evolution of the neutron star magnetic field might unfold as follows. (See Reisenegger 2009 for a more quantitative discussion.) When the neutron star is born, its internal temperature is high, $T\sim 10^{11}\mathrm{K}$. Its thermal energy, $E_T\sim 10^{52}(T/10^{11}\mathrm{K})^2\mathrm{erg}$, though much smaller than the gravitational binding energy, $|E_{grav}|\sim 10^{54}\mathrm{erg}$, is substantially larger than the magnetic energy, $E_{mag}\sim 10^{49}(B/10^{16}\mathrm{G})^2\mathrm{erg}$, in the early, relaxed, MHD equilibrium, even for the ultra-strong magnetar fields, $B\sim 10^{14-16}\mathrm{G}$. However, neutrino emission cools the star very quickly (much faster than the magnetic field can evolve), until it drops well below the ``equipartition'' line where $E_T=E_{mag}$, at which point the dissipation of even a small fraction of the magnetic energy can substantially feed back on the thermal evolution, essentially halting the cooling. For a strong magnetic field ($B\gtrsim 10^{16}\mathrm{G}$), this will happen in the high-temperature, strong-coupling regime ($T\gtrsim 10^9\mathrm{K}$), and the evolution of the magnetic field will be controlled by beta decays, whereas at lower $B$ the low-temperature, weak coupling regime is appropriate, and the evolution occurs through ambipolar diffusion, limited by neutron-charged particle collisions. Depending on field strength, the magnetic field in the crust might reorganize by breaking the latter (violently or causing plastic flow) or through Hall drift, or remain essentially unchanged, providing a fixed boundary condition to the evolution in the core. In any case, the magnetic feedback should leave $T$ essentially constant until the magnetic field has reached a new equilibrium state, compatible with the long-term, barotropic behavior of the liquid core matter. If (as I would conjecture) there are no stable magnetic equilibria in a barotropic, fluid sphere, then these long-lived magnetic equilibria in neutron stars will rely on being stabilized by the solid crust, and thus the typical field strength must be relatively weak, probably not reaching the magnetar range.

Clearly, the evolution of the magnetic field can be complex and will require numerical simulations to be sorted out in more detail, even in the absence of superfluidity and superconductivity, which I have ignored in the previous discussion. Some aspects of their effects have been considered by other authors, e.g., Glampedakis, Andersson, \& Samuelsson (2011).

\section{Conclusions}
\label{sec:conclusions}

The observed magnetic field strength on the surface of neutron stars appears to be roughly as expected from the flux of their progenitors (massive main sequence stars) and siblings (white dwarfs), although the neutron star birth is accompanied by violent processes that could alter it substantially. Soon after birth, it is likely to reach a stable, ideal-MHD equilibrium, with a poloidal and a toroidal component, which stabilize each other, aided by the compositional stratification of neutron star matter. The subsequent evolution relies on non-ideal-MHD processes such as Hall drift in the solid crust, and beta decays and ambipolar diffusion in the liquid core, all of which will lead to dissipation that temporarily halts the cooling of the neutron star, while the magnetic field re-arranges into a new equilibrium, which probably relies on shear forces in the crust that limit the field strength in this new, long-lived state.

\section{Acknowledgments}

I thank C. Armaza for preparing Fig.~\ref{fig:mustlie} and for technical assistance with the preparation of this manuscript, C. Petrovich for preparing Fig.~\ref{fig:BT}, and many colleagues and students for useful and stimulating discussions. This work was financially supported by FONDECYT Regular Grant 1110213, CONICYT International Collaboration Grant DFG-06, and the Basal Center for Astrophysics and Associated Technologies (PFB-06/2007).

\end{document}